\documentclass{article}
\usepackage{emulateapj,epsfig,eapjfig}

\slugcomment{DRAFT \today}

\shortauthors{BAKER ET AL.}
\shorttitle{Small-Scale Velocity Dispersion of the LCRS}

\newcommand\km{{\rm\ km}}
\newcommand\s{{\rm\ s}}
\newcommand\kms{{\rm\km\s^{-1}}}
\newcommand\kpc{{\rm\ kpc}}
\newcommand\Mpc{{\rm\ Mpc}}
\newcommand\hMpc{{h^{-1}\Mpc}}
\newcommand\hkpc{{h^{-1}\kpc}}
\newcommand\Msun{{\rm\ M}_\odot}

\newcommand\IRAS{{\sl IRAS\/}}
\newcommand\COBE{{\sl COBE\/}}

\newcommand\Dv{\Delta v}
\newcommand\Nex{N_{\mathrm{ex}}}
\newcommand\Nexi{N_{\mathrm{ex}, i}}
\newcommand\Nhi{N_{\mathrm{hi}}}
\newcommand\Nlo{N_{\mathrm{lo}}}
\newcommand\Dhi{D_{\mathrm{hi}}}
\newcommand\Dlo{D_{\mathrm{lo}}}
\newcommand\Ahi{A_{\mathrm{hi}}}
\newcommand\Alo{A_{\mathrm{lo}}}
\newcommand\rmin{r_{\mathrm{min}}}
\newcommand\rmax{r_{\mathrm{max}}}

\newcommand\rpmax{r_{p,\mathrm{max}}}
\newcommand\xibarbar{\overline{\overline{\xi}}}

\begin{document}

\title{The Galaxy-Weighted Small-Scale Velocity Dispersion \\
  of the Las Campanas Redshift Survey}

\author{Jonathan E.~Baker and Marc Davis}
\affil{Astronomy Department, University of California, Berkeley, CA 94720}
\email{\{jbaker, marc\}@astro.berkeley.edu}

\and

\author{Huan Lin\altaffilmark{1}}
\affil{Steward Observatory, University of Arizona,
  933 N. Cherry Avenue, Tucson, AZ 85721}
\email{hlin@as.arizona.edu}

\altaffiltext{1}{Hubble Fellow}

\begin{abstract} 
  The pair-weighted relative velocity dispersion of galaxies provides
  a measure of the thermal energy of fluctuations of the observed
  galaxy distribution, but the measure is difficult to interpret and
  is very sensitive to the existence of rare, rich clusters of
  galaxies.  Several alternative statistical procedures have recently
  been suggested to relieve these problems.  We apply a variant of the
  object-weighted statistical method of \markcite{DMW:97}Davis, Miller, \& White (1997) to the Las
  Campanas Redshift Survey (LCRS), which is the largest and deepest
  existing redshift survey that is nearly fully sampled.  The derived
  one-dimensional dispersion on scales $\sim 1\hMpc$ is quite low:
  $\sigma_1 = 126 \pm 10\kms$, with a modest decrease at larger
  scales.  The statistic is very stable; the six independent slices of
  the LCRS all yield consistent results.  We apply the same
  statistical procedure to halos in numerical simulations of an open
  cosmological model and flat models with and without a cosmological
  constant.  In contrast to the LCRS, all the models show a dispersion
  which increases for scales $> 1\hMpc$; it is uncertain whether this
  is a numerical artifact or a real physical effect.  The standard
  cluster-normalized Cold Dark Matter model with $\Omega_m=1$ as well
  as a tilted variant with $n=0.8$ yield dispersions substantially
  hotter than the LCRS value, while models with low matter density
  ($\Omega_m=0.3$) are broadly consistent with the LCRS data.  Using a
  filtered cosmic energy equation, we measure $\Omega_m\approx 0.2$,
  with small-scale bias factors $b=1.0$--$1.5$ for high-density models
  and $b=0.7$--$1.1$ for low-density models.
\end{abstract}


\section{Introduction}

The small-scale thermal energy of the observed galaxy distribution is
an important diagnostic for cosmological models.  For the past decade
the pair velocity dispersion $\sigma_{12}(r)$ \markcite{Davis:83}(Davis \& Peebles 1983) has
been the usual measure of this quantity (e.g., \markcite{Bean:83}Bean {et~al.} 1983;
\markcite{DeLap:88}de~Lapparent, Geller, \&  Huchra 1988; \markcite{Hale:89}Hale-Sutton {et~al.} 1989; \markcite{Mo:93}Mo, Jing, \& Borner 1993;
\markcite{Zurek:94,Fisher:94,Marzke:95,Brainerd:96}Zurek {et~al.} 1994; Fisher {et~al.} 1994; Marzke {et~al.} 1995; Brainerd {et~al.} 1996;
\markcite{Somerville:96,Landy:98,Jing:98}Somerville, Primack, \&  Nolthenius 1997; Landy, Szalay, \& Broadhurst 1998; Jing, Mo, \& Borner 1998).  But in spite of its
widespread application and the relative ease of its measurement within
large redshift surveys, the $\sigma_{12}(r)$ statistic has a number of
well-known deficiencies.  Chief among them is its pair-wise weighting,
which gives extreme influence to rare, rich clusters of galaxies
containing many close pairs with high velocity dispersion.

Alternative statistics to measure the thermal energy distribution have
been suggested by \markcite{Kepner:97}Kepner, Summers, \& Strauss (1997) and by \markcite{DMW:97}Davis, Miller, \& White (1997, hereafter
DMW).  The \markcite{Kepner:97}Kepner {et~al.} algorithm computes the
pair-weighted dispersion as a function of the local galaxy density;
this statistic demonstrates the heterogeneity of the environments of
the local galaxy distribution, but it must be computed in
volume-limited samples.  The $\sigma_1$ statistic described by DMW can
be estimated within a flux-limited catalog and is readily interpreted
in terms of a filtered version of the cosmic energy equation.  The
statistic is a measure of the rms one-dimensional velocity of
galaxies, with large-scale bulk flow motions filtered out.  DMW
applied this statistic to the UGC catalog of optical galaxies within
the Optical Redshift Survey \markcite{Santiago:95}(Santiago {et~al.} 1995), as well as the 1.2-Jy
\IRAS\ catalog \markcite{Fisher:95}(Fisher {et~al.} 1995).  They showed that $\Omega_m=1$
simulations were far too hot to match the observed dispersion.  Even
when compared with simulations in which the small-scale kinetic energy
had been artificially lowered by a factor of four, the observed
velocity distribution was colder than the simulated distribution.

However, the UGC catalog surveys a rather limited volume of the local
Universe, and the \IRAS\ catalog is quite dilute and under-samples
dense cluster regions.  It is therefore of considerable interest to
apply the DMW statistic to a larger, more representative redshift
survey such as the Las Campanas Redshift Survey
\markcite{Shectman:96}(LCRS; Shectman {et~al.} 1996), and to compare the results with $N$-body
simulations of cosmological models which are favored by current data.
This paper reports the application of this new statistic to the LCRS
and compares the result to a few simulations of flat and open
cosmological models.  In a future paper \markcite{Baker:99}(Baker, Davis, \& Ferreira 1999), we discuss a
wider variety of models, and we discuss in more detail the comparison
of the LCRS with $N$-body simulations and the potential applications
of $\sigma_1$ as a cosmological probe.

\section{Application of $\sigma_1$ to the LCRS}

The LCRS survey consists of 26,000 galaxies selected in a hybrid R
band.  The survey was conducted in six thin slices, each of size
$1\fdg5\times 80\arcdeg$ on the sky, with median redshift
$cz=30,000\kms$.  The redshift accuracy of the observations is
typically $\sigma_{\mathrm{err}} = 67\kms$ \markcite{Shectman:96}(Shectman {et~al.} 1996), which
is sufficient for measuring the thermal, small-scale velocity
dispersion.

For measurement of $\sigma_1$, we work with the subset of 19,306 LCRS
galaxies in the range $10,000 < cz < 45,000\kms$, and absolute
magnitude $-22.5 < M < -18.5$.  To estimate the random background of
the neighbors about each galaxy, we used a catalog of 268,000 randomly
distributed points with the same selection function as the LCRS
galaxies, including the restriction against pairs with angular
separation less than $55\arcsec$ caused by limitations of optical
fiber placement.  Since the six slices of the LCRS are spatially
separated by more than the projected separation used in the $\sigma_1$
statistic, the statistical procedure is applied to each slice
individually and the results are averaged.

\subsection{Method}

We now briefly describe our procedure, similar to that of DMW, for
determining $\sigma_1$.  For each galaxy $i$ in a slice of the survey,
we lay down a cylinder centered on the galaxy in redshift space.  Let
$r_p$ be the projected radius of the cylinder and $v_l$ its
half-length along the redshift direction.  For neighboring galaxies
$j$ within the cylinder, we construct the distribution $P_i(\Dv)$,
which counts the number of neighbors with redshift separation in a
redshift bin centered at $\Dv = v_j - v_i$.  The counts accumulated in
$P_i(\Dv)$ are weighted by the inverse selection function
$\phi_i/\phi_j$ (though equal weighting yields virtually identical
results).  We subtract from this distribution the background
distribution $B_i(\Dv)$, which counts the number of weighted neighbors
expected for an unclustered galaxy distribution.  We are interested in
the width of the overall distribution $D(\Dv)$ constructed by an
appropriately weighted sum over the $N_g$ galaxies:
\begin{equation}
  \label{eq:ddv}
  D(\Dv) = \frac{1}{N_g} 
  \sum_{i=1}^{N_g} w_i \left[ P_i(\Dv) - B_i(\Dv) \right],
\end{equation}
where the weight for galaxy $i$ is denoted by $w_i$.

In order to make the statistic object-weighted rather than
pair-weighted, we wish to normalize the distributions by the number of
neighbors $\Nex$ in excess of the random background, that is:
\begin{equation}
  \label{eq:whi}
  w_i^{-1} = \Nexi = \sum_{\Dv} \left[ P_i(\Dv) - B_i(\Dv) \right].
\end{equation}
This however presents a problem for galaxies which do not have enough
neighbors to ensure that the sum is positive.  DMW dealt with this
problem by deleting these objects from consideration, but under half
of the LCRS galaxies have at least one excess neighbor for
$r_p=1\hMpc$, and these galaxies are a biased sample because they
populate over-dense regions.  It is therefore desirable to modify the
statistic to include galaxies with fewer neighbors.

We achieve a more inclusive statistic by considering separately the
distributions of high- and low-density objects; that is, only galaxies
with $\Nex \geq 1$ are included in the sum for $\Dhi$, while only
galaxies with $\Nex < 1$ are included in the sum for $\Dlo$.  We then
weight the galaxies in the combined distribution according to
\begin{equation}
  \label{eq:wt}
  w_i = \left\{ 
      \begin{array}{l@{\qquad}l}
        \Ahi \Nexi^{-1} & \Nexi \geq 1 \\ \Alo & \Nexi < 1.
      \end{array}\right.
\end{equation}
Here $\Alo$ and $\Ahi$ are normalization constants for the two
distributions, chosen so that the distributions are weighted in
proportion to the number of objects included:
\begin{equation}
  \label{eq:ahi}
  \Ahi = \frac{\Nhi/N_g}{\sum_{\Dv} \left[ \Dhi(\Dv) - \Dhi(\infty) \right]},
\end{equation}
and similarly for $\Alo$.  Here $\Nhi$ and $\Nlo$ are the number of
galaxies with $\Nex \geq 1$ and $\Nex < 1$, respectively, thus $\Nhi +
\Nlo = N_g$.  The baselines $D(\infty)$ are estimated from the flat
tails of the distributions within $500\kms$ of $\Dv = \pm v_l$.  With
this normalization the final distribution obeys $\sum_{\Dv} D(\Dv) =
1$.  Note that scaling $\Dhi$ and $\Dlo$  by the constants $A$ does not
affect the derived widths for these distributions; rather, it merely
alters the weighting of the two in the combined distribution.

This procedure, in contrast to that of DMW, allows us to include all
of the available data, yielding an unbiased, object-weighted measure
of the thermal energy of the galaxy distribution.  It is the
object-weighting which differentiates our procedure from the more
traditional measure of the pair dispersion $\sigma_{12}(r)$; all
galaxies (not pairs) are assigned equal weight in our statistic
$\sigma_1$.

We measure the width of the distribution $D(\Dv)$ using the
convolution procedure outlined by DMW (equation 18), in which a
velocity broadening function $f(v)$ is convolved with the two-point
correlation function $\xi(r)$ to produce a model $M(\Dv) =
\overline{\xi}_{r_p} \ast f$ for $D(\Dv)$:
\begin{equation}
  \label{eq:model}
  M(\Dv) = \int\limits_0^{r_p} dr\, 2\pi r \int\limits_{-\infty}^\infty  
  dy\, \xi(\sqrt{r^2 + y^2})\, f(\Dv - y).
\end{equation}
The two-point correlation function of the LCRS is well-approximated by
$\xi(r) = (r_0/r)^\gamma$, with $r_0=5\hMpc$ and $\gamma=1.8$
\markcite{Jing:98}(Jing {et~al.} 1998), while for the $N$-body simulations we use the
cylindrically averaged mass correlation function
$\overline{\xi}_{r_p}(\Dv)$ measured directly from the particle
distribution.  We find that an exponential broadening function
(see \markcite{1996ApJ...467...19D,Sheth:96}{Diaferio} \& {Geller} 1996; Sheth 1996, \markcite{Jusz:98}Juszkiewicz, Fisher, \& Szapudi 1998)
\begin{equation}
  \label{eq:fv}
  f(v) = \frac{1}{\sigma_1} \exp\left(-\frac{|v|}{\sigma_1}\right)
\end{equation}
provides a much better fit to the LCRS data and all $N$-body models
than does a Gaussian.  Here we have defined the width $\sigma_1$ so
that it is a measure of the rms velocity of individual galaxies in one
dimension (with bulk motions on scales $\gtrsim 1\hMpc$ filtered out).
The (object-weighted) rms difference in velocity between any two
galaxies is then $\sigma_1\sqrt{2}$ (DMW call this quantity, which is
equal to the rms dispersion of the distribution $f$, the ``intrinsic''
dispersion $\sigma_I$; we will work exclusively with $\sigma_1$ to
avoid confusion).  
\eapjfig{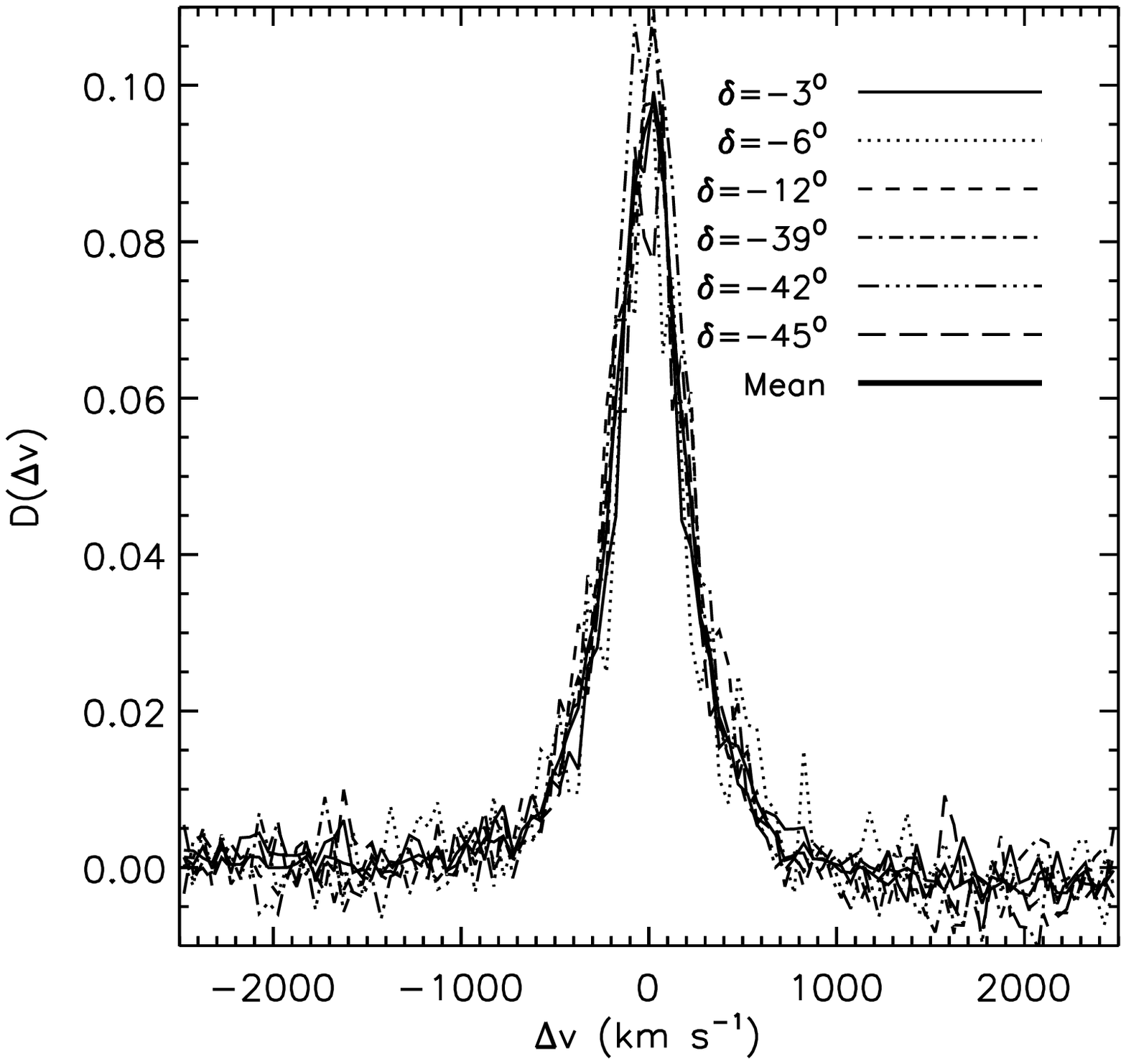}{Galaxy-weighted velocity distribution $D(\Dv)$ for 
  the six LCRS slices.\label{fig:slices}}{0.15in}
The three-dimensional dispersions are larger by an additional factor
$\sqrt{3}$.

We perform a nonlinear $\chi^2$-minimization fit to determine the
width $\sigma_1$ and amplitude of the model $M(\Dv)$.  Before fitting,
we convolve the model with a Gaussian of rms $\sigma_{\mathrm{err}}
\sqrt{2} = 95\kms$ to account for the LCRS redshift measurement
uncertainties; the factor of $\sqrt{2}$ converts from the measurement
uncertainty for individual redshifts to the uncertainty for redshift
differences, which are accumulated in $D(\Delta v)$.  We also include
baseline terms in the model which are constant and linear in $\Dv$,
for a total of four fit parameters.  The linear term is necessary for
the LCRS because for simplicity we define ``cylinders'' in redshift
space based on projected angular separation on the sky.  This leads to
a small gradient in the measured distribution function $D(\Dv)$
because the ``cylinders'' are in fact conic sections, but the term is
quite small because the length of the cylinders, $2v_l$, is much
smaller than the typical redshift of galaxies in the survey.  Although
the gradient term has a negligible effect on the derived width, it
does improve significantly the quality of the $\chi^2$ fit.

\subsection{Results for the LCRS\label{sec:lcrs}}

We have used the six independent slices of the LCRS to estimate the
errors in $D(\Dv)$ as a function of $\Dv$ in computing $\chi^2$.
However, we expect that the bins may be correlated due to sample
variance; the fitting procedure is therefore not strictly legitimate,
but the consistency of the results for the widths of the individual
slices serves as a check on the degree to which sample variance
affects the result.  We also expect $\chi^2_\nu > 1$ if the
exponential broadening function of width $\sigma_1$ (assumed
independent of $r$) provides an inadequate description of the
small-scale velocities.

\begin{table*}
  \begin{center}
    \begin{tabular}{c|r|ccc|c|r}
      Slice & Dec. & 
      \multicolumn{3}{c|}{$\sigma_1$ (km s$^{-1}$)} & 
      $\Nhi/N_g$ & $N_g$ \\
      & & All & $\Nex \geq 1$ & $\Nex < 1$ & & \\
      \tableline
      1 & $-3^\circ$  &  96 & 184 &  53 & 0.44 & 3540 \\
      2 & $-6^\circ$  & 103 & 255 &  73 & 0.34 & 2067 \\
      3 & $-12^\circ$ & 163 & 273 & 129 & 0.44 & 3754 \\
      4 & $-39^\circ$ & 117 & 181 &  94 & 0.41 & 3265 \\
      5 & $-42^\circ$ & 136 & 178 & 112 & 0.44 & 3503 \\
      6 & $-45^\circ$ & 142 & 171 & 131 & 0.43 & 3177 \\[0.5ex]
   Mean &             & {\bf $126 \pm 10$} & $207 \pm 18$ & 
                        $99 \pm 13$ & 0.42 & 3218 \\[0.5ex]
   1--6 &             & 136 & 208 & 101 & 0.42 & 19306 \\
    \end{tabular}
  \end{center}
  \caption{The $\sigma_1$ statistic for the six LCRS slices.  The
    second to last line gives the mean and standard deviation of 
    the mean of the independent slices, while the last line is the 
    result of a fit to the entire dataset.  We list $\sigma_1$ for the 
    combined dataset and for the subsets of galaxies with $\Nex \geq 1$ and
    $\Nex < 1$.  $\Nhi/N_g$ is the fraction of galaxies with $\Nex
    \geq 1$, and $N_g$ is the total number of galaxies.
    \label{tab:slices}}
\end{table*}

The $D(\Dv)$ distributions for the six independent LCRS slices are
plotted in Figure~\ref{fig:slices}, and Table~\ref{tab:slices} lists
the derived widths.  The second to last line gives the mean and
standard deviation of the mean for separate fits to the six slices,
while the last line is the result of a single fit to the combined
distribution of all galaxies.  Note that 
the dispersion measured for objects with excess neighbors ($\Nex \geq
1$) is clearly higher than that measured for objects with fewer
neighbors.  This behavior is expected because objects with more
neighbors are found in regions of higher density, which tend to be
hotter.

The fit to the LCRS $D(\Dv)$, shown in Figure~\ref{fig:lcrsfit}, is
quite good, with $\chi_\nu^2 = 117/96 = 1.22$; the probability of
$\chi^2$ exceeding this value is $1-P(\chi^2|\nu) = 7\%$.  The
best-fitting Gaussian $f(v)$ is much worse, with $\chi_\nu^2 = 1.84$
and $1-P(\chi^2|\nu) = 10^{-6}$.

Based on the mean of the six slices we adopt $\sigma_1 = 126 \pm 10
\kms$.  This value has been computed for $r_p=1\hMpc$ and
$v_l=2500\kms$.  The results are quite insensitive to cylinder length,
ranging only from $117 \pm 14\kms$ at $v_l=1500\kms$ to $132 \pm
13\kms$ at $v_l=3500\kms$.  Our chosen value $v_l=2500\kms$ is large
enough to allow a clean measure of the tails of the distribution
without significant non-linearities in the baseline gradient due to
variations in the selection function.

\begin{table*}[b]
  \begin{center}
    \begin{tabular}{ccc}
      $r_p (\hMpc)$ & $\sigma_1 (\mathrm{km}\s^{-1})$ & $\Nhi/N_g$ \\
      \tableline
      0.5 & $136 \pm 10$ & 0.23 \\
      1   & $126 \pm 10$ & 0.42 \\
      1.5 & $107 \pm 8$  & 0.55 \\
      2   & $96 \pm 12$  & 0.63 \\
      2.5 & $99 \pm 13$  & 0.68 \\
    \end{tabular}
    \caption{LCRS dispersion $\sigma_1$ as a function of limiting
      projected radius $r_p$, and fraction of galaxies with excess
      neighbors.}
    \label{tab:rp}
  \end{center}
\end{table*}

A modest decrease in $\sigma_1$ is evident as $r_p$ is increased above
$r_p=1\hMpc$ (see Table~\ref{tab:rp}).  Although the $D(\Dv)$
distributions are very insensitive to $r_p$, the averaged correlation
function $\overline{\xi}_{r_p}(\Dv)$ becomes broader as $r_p$ increases.
As a result, smaller values of $r_p$ provide a cleaner measure of the
true (real-space) velocity broadening on small scales, but decreasing
$r_p$ below $1\hMpc$ reduces the signal-to-noise, as most galaxies
have too few neighbors.  The background subtraction also becomes
cleaner as $r_p$ is reduced.

Note that for the larger value $r_p=2\hMpc$ used 
\eapjfig{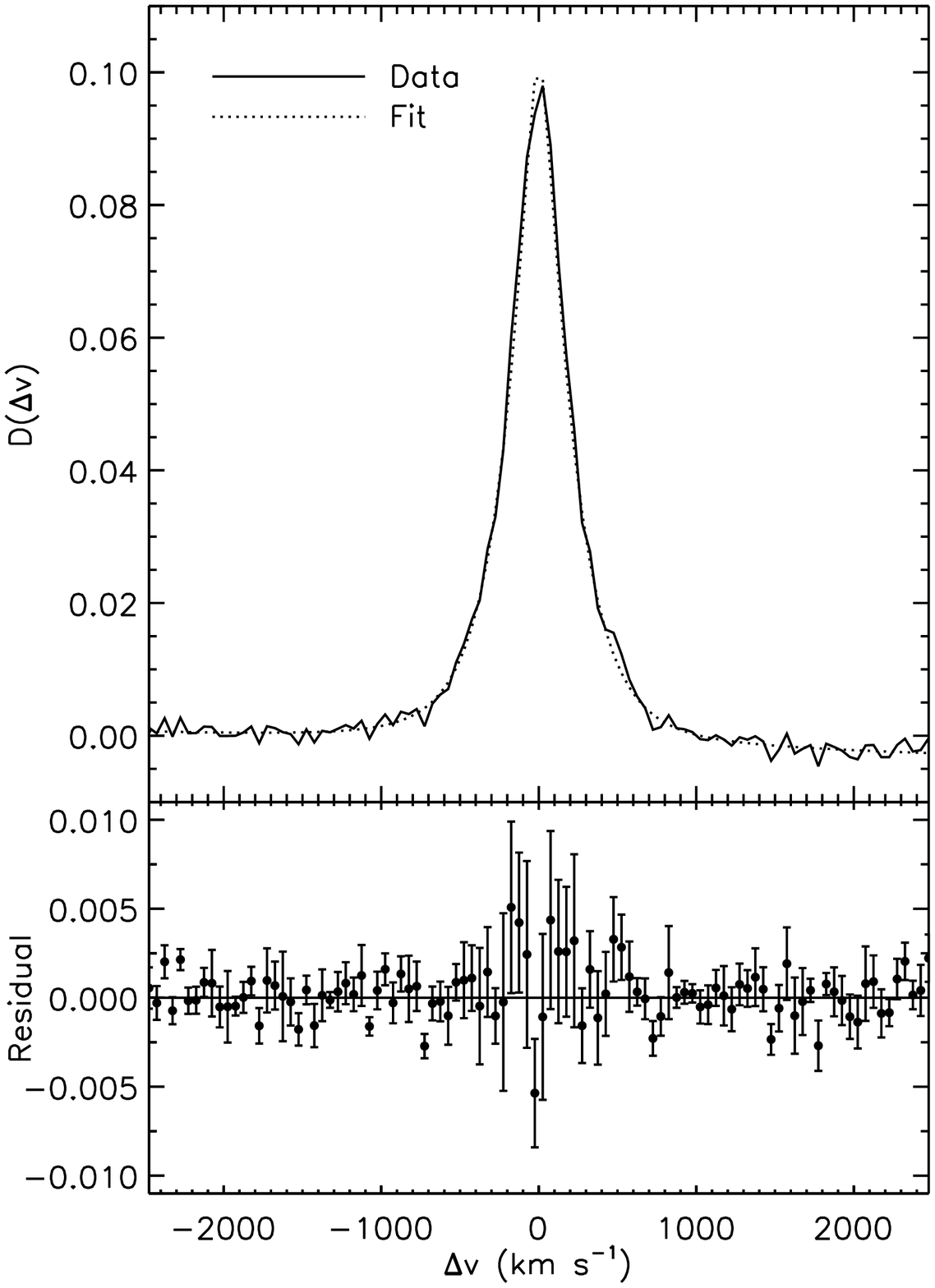}{Velocity distribution $D(\Dv)$ and fit for the combined 
  LCRS data (upper panel), and residuals for the fit with errors 
  estimated from the standard deviation of the six slices (lower 
  panel; note the change in vertical scale).
  \label{fig:lcrsfit}}{0in}
by DMW, our result is $\sigma_1 = 114\pm 10\kms$.  If, as in the DMW
analysis, we do not account for broadening due to redshift measurement
errors, the result increases to $\sigma_1 = 136\pm 9\kms$.  Since the
two surveys have comparable redshift uncertainties, our LCRS result is
perfectly consistent with the value $\sigma_1 = 130\pm15\kms$ which
DMW derived for the much smaller UGC catalog.

\section{Comparison to $N$-body models}

We have completed a suite of $N$-body simulations designed to predict
the small-scale velocity dispersion in a variety of cosmological
models.  Here we discuss the results of a few of these models: the
``standard'' Cold Dark Matter (SCDM) model and a tilted variant
(TCDM), a model with a cosmological constant $\Lambda$ (LCDM), and an
open model (OCDM).  The cosmological parameters for these models are
listed in Table~\ref{tab:cosmo}.  All models are approximately
normalized to the present-day abundance of clusters; the LCDM and TCDM
models additionally satisfy the \COBE\ normalization.  The SCDM model
is known to fail a number of cosmological tests and is included for
historical reasons, and only LCDM is fully consistent with current
limits from high-redshift supernovae \markcite{Perlmutter:99}({Perlmutter} {et~al.} 1999).  We note
that on the scales relevant for our simulations, the TCDM power
spectrum is indistinguishable from a $\tau$CDM spectrum with shape
parameter $\Gamma=0.2$.  A broader range of models and a more detailed
discussion of the simulations may be found elsewhere \markcite{Baker:99}(Baker {et~al.} 1999).

\begin{table*}
  \begin{center}
    \begin{tabular}{cccccc}
      Model & $\Omega_m$ & $\Omega_\Lambda$ & $n$ & $h$ & $\sigma_8$
      \\
      \tableline
      SCDM & 1   & 0   & 1   & 0.5 & 0.7 \\
      TCDM & 1   & 0   & 0.8 & 0.5 & 0.7 \\
      LCDM & 0.3 & 0.7 & 1   & 0.7 & 1 \\
      OCDM & 0.3 & 0   & 1   & 0.7 & 1 \\
    \end{tabular}
    \caption{Cosmological parameters for $N$-body models: matter
      density $\Omega_m$, cosmological constant $\Omega_\Lambda$, tilt 
      $n$ where $P(k)\propto k^n$, Hubble constant $H_0 =
      h/100\kms\Mpc^{-1}$, and rms mass fluctuation $\sigma_8$ in
      spheres of radius $8\hMpc$.}
    \label{tab:cosmo}
  \end{center}
\end{table*}

Initial power spectra were obtained using the CMBFAST code
\markcite{Seljak:96,CMBFAST}(Seljak \& Zaldarriaga 1996, 1998).  The simulations were evolved on a $128^3$
mesh using a P$^3$M code \markcite{Brieu:95}(Brieu, Summers, \& Ostriker 1995) in which short-range forces
are computed using a special purpose GRAPE-3AF board
\markcite{Okumura:1993}(Okumura {et~al.} 1993).  We chose a box of size $L=50\hMpc$ to match the
length of the LCRS cylinders; with $N_p=64^3$ particles this gives a
mass resolution of $1.3\times 10^{11}\Omega_m h^{-1}\Msun$, where
$h=H_0/100\kms\Mpc^{-1}$.  A Plummer force softening
$\epsilon=50\hkpc$ was used.  The simulations were started at
redshifts $z_i=15$ (for $\Omega_m=1$) or $z_i=19$ (for $\Omega_m=0.3$)
and evolved to $z=0$ in 1500 time-steps using $p=a^2$ as the
integration variable.

\begin{figure*}[b]
  \begin{center}
    \epsfig{width=6.2in,file=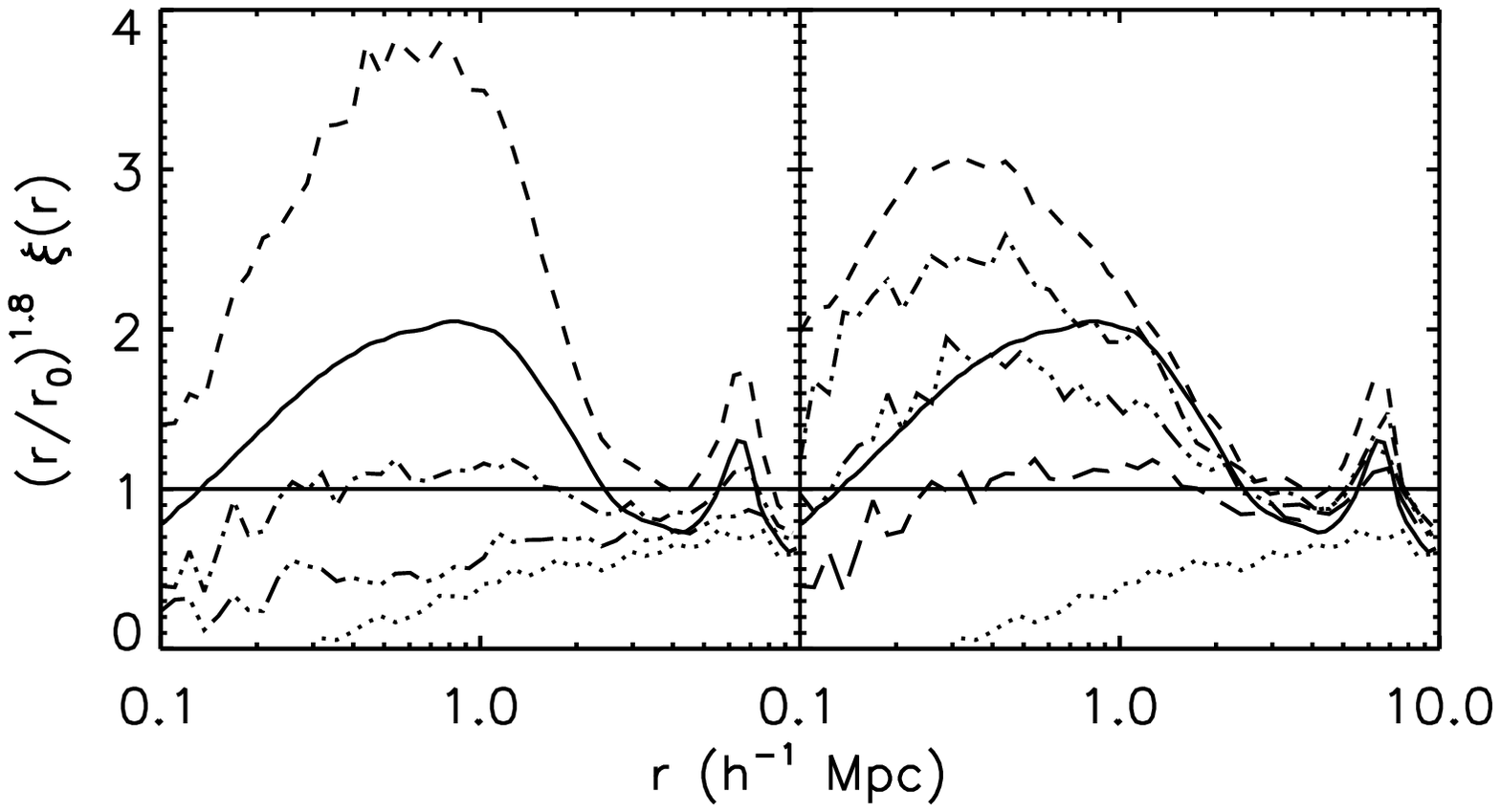}
    \caption{Two-point correlation functions multiplied by
      $(r/r_0)^{1.8}$ for LCDM simulation halos for a range of $N_s$ and
      $\alpha$.  Here $r_0 = 5.1\hMpc$ and the LCRS $\xi(r)$ appears as
      a solid horizontal line.  In both plots, the curve at bottom
      (dotted line) shows $N_s=\infty$ (no halo splitting), and the
      solid curve shows the mass correlation function.  In the plot at
      left, $N_s$ is held fixed at 80, while $\alpha$ takes on the
      values 0, 0.25, 0.5 (top to bottom).  In the plot at right,
      $\alpha$ is held fixed at 0.25, while $N_s$ takes on the values
      10, 20, 40, 80 (top to bottom).\label{fig:nsa}} 
  \end{center}
\end{figure*}

The simulations are converted to ``redshift'' space by adding the
velocities $v_i$ along one of the three coordinates $i$ to the
positions $x_i$: $x_i \rightarrow x_i + v_i/H$, where $H$ is the
Hubble constant.  Periodic boundary conditions are applied at the box
edges.  We then apply exactly the same statistical procedure for
determining $\sigma_1$ as for the LCRS, except that the selection
function is now unity.

\subsection{Tests of $\sigma_1$ Measurements}

We have used our $N$-body simulations to perform a number of checks on
the robustness of our method for determining the small-scale velocity
dispersion.  One test is to ask how well our model is able to account
for the redshift measurement uncertainties in the LCRS\@.  To simulate
these uncertainties, we added Gaussian random velocities of rms
$\sigma_{\mathrm{err}}$ along the ``redshift'' coordinate in the
simulations.  We then make two determinations of $\sigma_1$, which
should ideally be equal.  In one determination, the random velocities
have been added and we perform an extra Gaussian convolution in the
model to account for them.  In the other, no random velocities are
added and no Gaussian convolution is necessary.  We find that the two
widths agree quite well, to within $10\kms$ over the range of interest
for $\sigma_1$ (100--300$\kms$).  The agreement improves as $\sigma_1$
increases and the uncertainties contribute relatively less to the
width of the observed velocity distribution.

A second test of the method is to compare velocity widths measured in
real space with those measured in cylinders in redshift space.  For
this test, we replace the velocities of the simulation particles with
velocities drawn from a random exponential distribution of a given rms
$\sigma$.  It is straightforward to show that the velocity
distribution appropriate for the difference distribution $D(\Dv)$ is
then
\begin{equation}
  \label{eq:fvd}
  f(v) = \frac{1}{2\sigma^2} \left(|v| +
    \frac{\sigma}{\sqrt{2}}\right) \exp{\left( -\sqrt{2}
      \frac{|v|}{\sigma} \right)}.
\end{equation}
Using this form in the redshift-space model (Equation~\ref{eq:model}),
we find that our procedure recovers the true velocity dispersion with
an accuracy better than 10\% for $\sigma_1$ in the range
100--300$\kms$.

Finally, we can test the extent to which our measurement of $\sigma_1$
in the long redshift-space cylinders is contaminated by motions on
scales larger than $1\hMpc$.  First we construct distributions
analogous to $D(\Dv)$, but measured in real space, with neighbors
drawn from spheres of radius $1\hMpc$ in the simulations.  These are
compared to distributions with neighbors drawn from the long
cylinders, also measured in real space.  The widths of these
distributions agree to within 1\%, and we conclude that the
contamination from large scales is negligible.

\begin{figure*}[t]
  \begin{center}
    \epsfig{width=6.2in,file=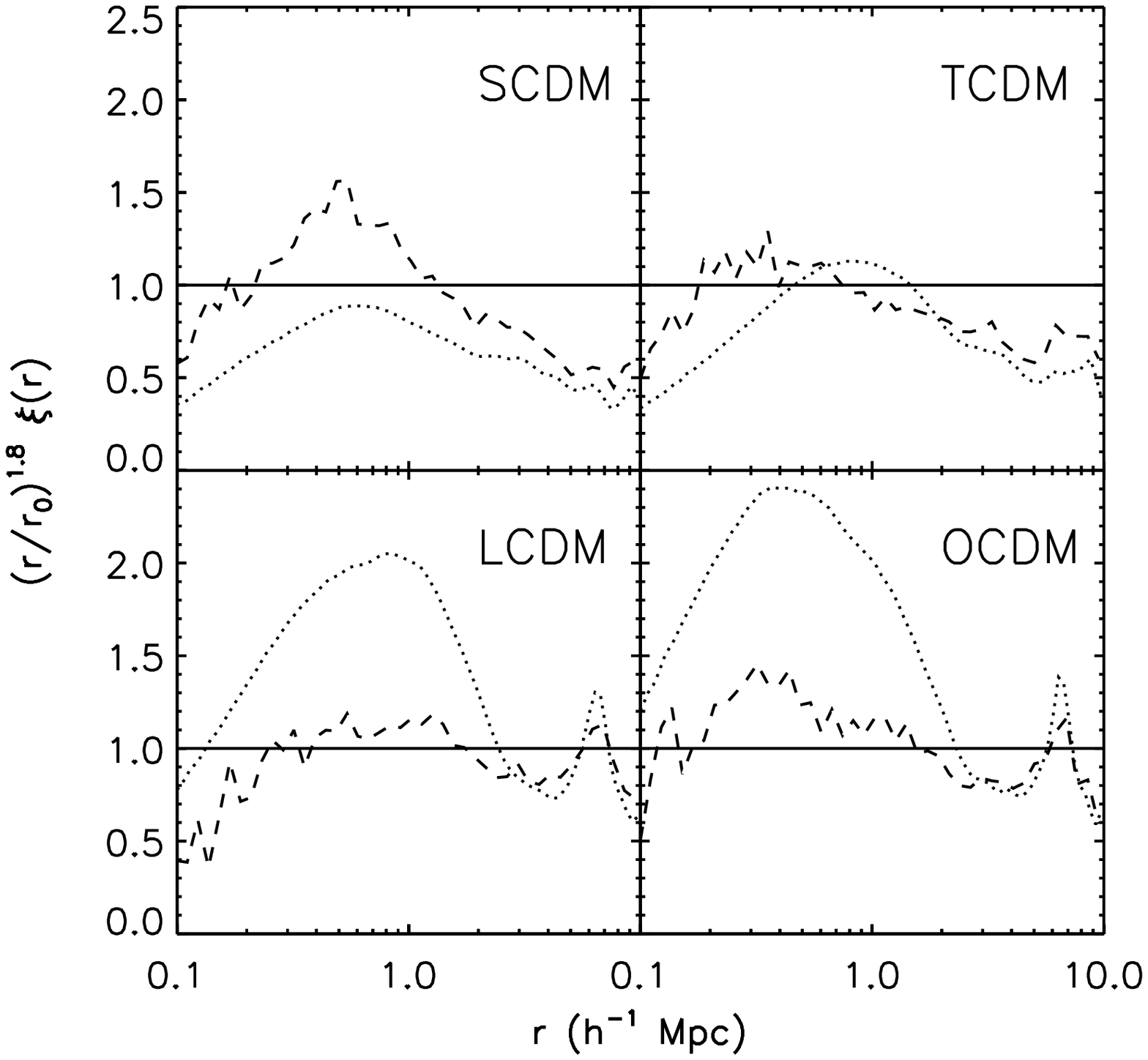}
    \caption{Two-point correlation functions for the mass (dotted
      lines), halos (dashed), and LCRS (solid), plotted as in
      Fig.~\ref{fig:nsa}.  Halos were selected using the parameters
      listed in Table~\ref{tab:sig1nbody}.}
    \label{fig:xi}
  \end{center}
\end{figure*}

\subsection{Selection of Galaxies from the Mass Distribution}

We can easily compute $\sigma_1$ for particles in the simulations, but
the observed small-scale dispersion of galaxies, which correspond in
some way to halos in the simulations, will in general differ from that
of the mass.  The internal velocity dispersions of galaxies are not
included in the observed statistic; moreover, the galaxy population
may be a biased tracer.  In order to test whether our simulations can
reproduce the LCRS result for $\sigma_1$, it is therefore important to
identify ``galaxies'' within the $N$-body simulations.  Unfortunately
the process of galaxy formation includes baryonic physics on a wide
range of scales not probed by our dark-matter only simulations.  For
the present work, we define galaxies using a simple phenomenological
model which we expect to yield similar results to those of larger
gas-dynamical simulations.

We first apply the standard friends-of-friends \markcite{DEFW:85}(FOF; Davis {et~al.} 1985)
algorithm to the simulations, with a linking length of 0.2 mesh
cells and a minimum group size $N\geq 10$, corresponding to halos with
mass $M\gtrsim 10^{12}\Omega_m h^{-1}\Msun$.  We have also considered
the HOP method \markcite{Eisenstein:98}(Eisenstein \& Hut 1998) for defining halos, but we obtain
similar results for reasonable parameter choices and do not discuss
them here.

Our limited resolution and the nature of the FOF algorithm lead to a
serious and well-known over-merging problem, in which a large cluster
containing many galaxies will be identified as a single halo.  This
drastically lowers the small-scale velocity dispersion because the
motions of galaxies within clusters are neglected.  To remedy this
situation, we split halos with more than $N_s$ particles 
by randomly selecting particles from within the halos and identifying
these particles as galaxies.  Halos identified in this way will again
include the internal motions of galaxies, but as the splitting is only
applied to large, hot halos ($N_s\gg 10$), we expect these internal
motions to have a negligible effect on our result.  Small halos with
fewer than $N_s$ particles are taken to be individual galaxies.

\begin{table*}
  \begin{center}
    \begin{tabular}{cccccc}
      Model & $N_s$ & $\alpha$ & $N_{\mathrm{halos}}$ &
      \multicolumn{2}{c}{$\sigma_1$ ($\mathrm{km}\s^{-1}$)} \\
      & & & & Mass & Halos \\
      \tableline
      SCDM & 80 & 0.00 & 2624 & 310 & 269 \\
      TCDM & 40 & 0.25 & 2351 & 320 & 293 \\
      LCDM & 80 & 0.25 & 1901 & 188 & 143 \\
      OCDM & 80 & 0.20 & 2128 & 197 & 158 \\
    \end{tabular}
    \caption{Velocity width $\sigma_1$ for the $N$-body simulations.
      Results are listed for all particles and for the
      $N_{\mathrm{halos}}$ halos identified using our splitting
      procedure with parameters $N_s$ and $\alpha$.}
    \label{tab:sig1nbody}
  \end{center}
\end{table*}

For comparison with the LCRS, we wish to choose a set of halos which
resemble the LCRS galaxies as closely as possible.  Some $N$-body
models yield a correlation function $\xi(r)$ which is too steep, and
it is therefore advantageous to select halos which are anti-biased on
small scales \markcite{Jing:98}(Jing {et~al.} 1998).  We accomplish this through our
halo-splitting procedure by drawing random particles with a
probability $p$ which has a power-law dependence on the number of
particles $N$ in the parent halo: $p(N) = N_s^{\alpha-1} N^{-\alpha}$,
with $\alpha>0$.  The number of galaxies per unit halo mass then falls
as $N^{-\alpha}$ for large halos.  We choose parameters $N_s$ and
$\alpha$ which simultaneously mimic the power-law shape of the LCRS
correlation function and produce approximately the correct number
density of galaxies, $n\approx 0.02 h^3\Mpc^{-3}$, implying 2500
galaxies per simulation volume.  Increasing $\alpha$ tends to flatten
the correlation function on small scales and yields fewer halos;
increasing $N_s$ at fixed $\alpha$ tends to lower the correlation
amplitude and also yields fewer halos.  This behavior is illustrated
in Figure~\ref{fig:nsa} for the LCDM model.

Figure~\ref{fig:xi} shows the correlation functions for our selected
halos in each of the models.  In the low-density models, we are able
to select halos which match the LCRS $\xi(r)$ quite well.  The
normalization of the high-density models is such that $\xi(r)$ always
falls below the LCRS power law on large scales.  The TCDM halos match
well at $r\lesssim 2\hMpc$.  In the SCDM model, we are unable to
reproduce exactly the shape of $\xi(r)$ without falling too far below
the LCRS amplitude and producing too few halos.  However, the
differences in $J_2$ (see \S\ref{cosmic}) computed from these
correlation functions show that this mismatch should affect our
estimate of $\Omega_m$ by at most 30\%.

\subsection{Results for $\sigma_1$}

The results for $\sigma_1$ for our four cosmological models are listed
in Table~\ref{tab:sig1nbody}.  We see that the mass in the two
$\Omega_m=1$ models is far too hot on $1\hMpc$ scales, with $\sigma_1$
well over twice the LCRS value.  The spectral tilt of the TCDM model
has very little effect on the small-scale velocities, as the result is
nearly identical to the SCDM result.  The mass in the low-$\Omega_m$
models, on the other hand, is also hotter than the LCRS, but only by a
factor of about 1.5.  

The halos in the simulations are somewhat cooler than the mass, with
small-scale dispersions lower by factors in the range 0.7--0.9.  The
LCDM halos come closest to the LCRS value; at $143\kms$, they are only
marginally ($1.7\sigma$) hotter than the LCRS\@.  The open model
produces velocity dispersions slightly higher than the LCDM model,
while the halos in the $\Omega_m=1$ models are again much hotter than
the LCRS data.

\begin{figure*}[t]
  \begin{center}
    \epsfig{width=6.2in,file=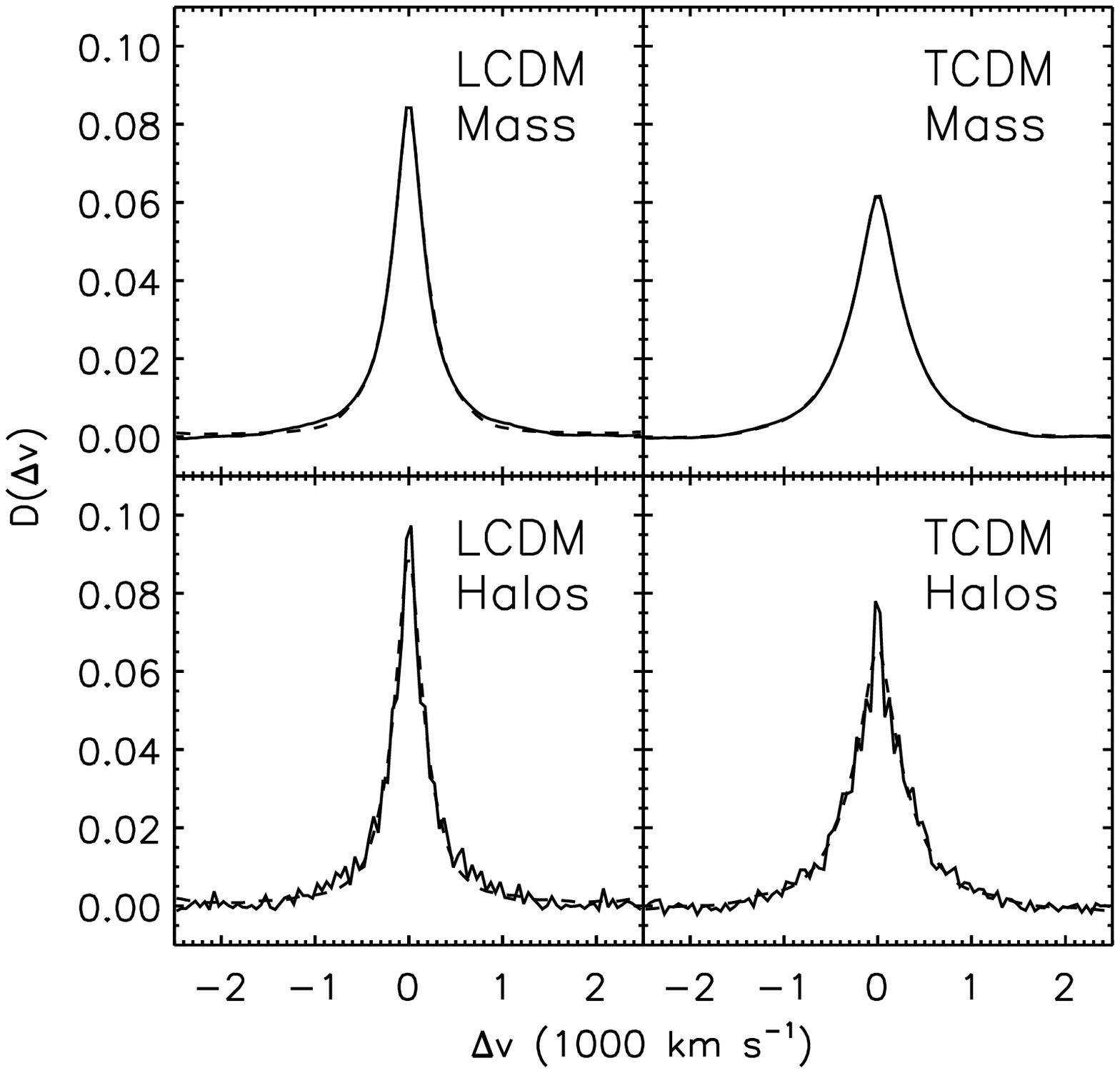}
    \caption{Velocity distributions (solid) and fits (dashed) for two 
      representative $N$-body models.  In each plot the fit is shown as
      a dashed line and is generally invisible because it is obscured
      by the measured distribution.  Top and bottom panels show the
      distributions for all particles and for halos, respectively.}
    \label{fig:nfit}
  \end{center}
\end{figure*}

Figure~\ref{fig:nfit} shows that the exponential $f(v)$ provides an
excellent fit to the velocity distributions measured in the
simulations in redshift space.  We show distributions for the $N$-body
mass particles and for the halos.  The halo distributions are noisier
because there are many fewer halos than mass particles in the
simulation volumes.  The distributions for the SCDM and OCDM models
are nearly indistinguishable from the TCDM and LCDM distributions,
respectively, and are not shown.

We have also computed $\sigma_1$ for galaxies drawn using more
sophisticated semi-analytic techniques from a large Virgo simulation
\markcite{Benson:99}({Benson} {et~al.} 1999) of the LCDM model.  This simulation has a mass
resolution better than ours by about a factor of two, and the box
length is nearly three times as large.  The result is $126\kms$, only
slightly lower than our value of $143\kms$.  This suggests that our
procedure for defining galaxies is reasonable.  The Virgo result
exactly matches the LCRS dispersion, which suggests that the
small-scale velocity dispersion predicted by the $\Omega_m=0.3$ flat
model is in fact perfectly consistent with the observational data.
Further details of this comparison will be presented in a future work
\markcite{Baker:99}(Baker {et~al.} 1999).

As noted in \S\ref{sec:lcrs}, the LCRS velocity width decreases
somewhat as the limiting radius $\rpmax$ is increased.  In
Figure~\ref{fig:scale}, we show this scale dependence measured in
independent cylindrical shells of width $1\hMpc$, where the limits on
the radial integration in the model (Equation~\ref{eq:model}) have
been adjusted appropriately.  Although the measured LCRS $D(\Delta v)$
shows little scale dependence, the integrated correlation function
broadens with scale, leading to a smaller measured velocity width.

None of the $N$-body models, however, are able to reproduce the scale
dependence observed in the LCRS\@.  The halos drawn from the Virgo
simulation, which show very little scale dependence, come closest,
while the other models tend to show an increase in velocity dispersion
with scale.  Only the LCDM model is shown in Figure~\ref{fig:scale},
but we find similar discrepancies for the other models as well.
Although the $\Omega_m=0.3$ LCDM model produces a reasonable match to
the velocity dispersion on very small scales, all of the models seem
unable to reproduce the observed coldness of the velocities on
intermediate scales $\sim 1$--$3\hMpc$.  At present it is unclear
whether this discrepancy is due to problems with the galaxy selection
procedure, the resolution of the simulations, or a more fundamental
flaw in the cosmological models.

\subsection{Filtered Cosmic Energy Equation\label{cosmic}}

The $\sigma_1$ statistic is ideally suited for the application of the
cosmic energy (Layzer-Irvine) equation filtered on small scales.  As
shown by DMW, we expect $\sigma_1^2 \propto \Omega_m J_{2,m}$ in the
absence of velocity bias, where
\begin{equation}
  \label{eq:j2}
  J_2 = \int_{\rmin}^{\rmax} dr\, r\xi(r).
\end{equation}
The subscript $m$ means that $J_2$ is computed from $\xi_m(r)$, the
correlation function for the underlying mass.  We can write this in
terms of the measured $\xi(r)$ of an observed sample $j$ by defining
an effective bias $b_j^2 = J_{2,j}/J_{2,m}$.  If we then compare
$\sigma_{1,j}$ measured for sample $j$ with $\sigma_{1,N}$ measured
for the underlying mass in an $N$-body simulation with mass density
parameter $\Omega_N$, we can measure the parameter
\begin{equation}
  \label{eq:omb2}
  \Omega_m/b_j^2 = 
  \left( \frac{\sigma_{1,j}}{\sigma_{1,N}} \right)^2
  \left( \frac{J_{2,N}}{J_{2,j}} \right) \Omega_N.
\end{equation}
If in addition we can choose a sample of $N$-body halos which matches
the correlation function of the sample $j$, then we have a direct
measure of $\Omega_m$:
\begin{equation}
  \label{eq:om}
  \Omega_m = (\sigma_{1,j}/\sigma_{1,N})^2 \Omega_N,
\end{equation}
where $\sigma_{1,N}$ is now measured for the $N$-body halos rather
than the underlying mass.

\begin{table*}
  \begin{center}
    \begin{tabular}{cccc}
      Model & $\Omega_m$ & $\Omega_m/b^2$ & $b$ \\
      \tableline
      SCDM & 0.22$\pm$0.03 & 0.10--0.14 & 1.2--1.5 \\
      TCDM & 0.18$\pm$0.03 & 0.12--0.15 & 1.0--1.3 \\
      LCDM & 0.23$\pm$0.04 & 0.20--0.27 & 0.8--1.1 \\
      OCDM & 0.19$\pm$0.03 & 0.20--0.29 & 0.7--1.0 \\
    \end{tabular}
    \caption{Density parameter and small-scale bias derived from the
      cosmic energy equation and the LCRS dispersion.}
    \label{tab:omega}
  \end{center}
\end{table*}

The results of combining the LCRS dispersion $\sigma_1 = 126\pm
10\kms$ with our four cosmological $N$-body models are listed in
Table~\ref{tab:omega}.  Based on the halos in each of the four
simulations, we derive consistent values $\Omega_m\approx 0.2$.  Note
that the errors listed on $\Omega_m$ are 1-$\sigma$ uncertainties
derived solely from the LCRS $\sigma_1$ result; they do not include
any systematic errors in the model results.  The fact that we derive
similar values of $\Omega_m$ from each of the different models is
an important consistency check, and gives us confidence that our
method is indeed a sensitive probe of the matter density.

Table~\ref{tab:omega} also lists the values of $\Omega_m/b^2$ derived
by comparing the LCRS dispersion with the dispersion of the $N$-body
mass.  The integral $J_2$ converges rather slowly, and its value is
quite sensitive to the integration limits $\rmin$ and $\rmax$.  A
reasonable lower limit is $\rmin=0.1\hMpc$, which eliminates from the
analysis the internal velocity dispersion of typical galaxies and
includes only the dispersion of galaxies moving relative to each
other.  We might also take $\rmax$ to be slightly larger than
$1\hMpc$, since the length of the redshift-space cylinders means that
there will be some contribution to $\sigma_1$ from larger scales
(although we have measured this effect in the simulations and have
found that it is very small).  The ranges shown for $\Omega_m/b^2$
were obtained by allowing $\rmin$ and $\rmax$ to vary over the ranges
0.05--0.2 and 1--5 $\hMpc$, respectively.  Our results for the
high-density models are consistent with the value $\Omega_m/b^2 =
0.14\pm 0.05$ found by DMW, who only considered an $\Omega_m=1$ model.

The parameter $\Omega_m/b^2$ is approximately equal to $\beta^2$,
where $\beta \approx \Omega_m^{0.6}/b$ is the parameter measured by
large-scale flow analyses.  We find $\beta\approx 0.3$--$0.4$ for the
two high-density models, and $\beta\approx 0.45$--$0.55$ for the two
low-density models.
These ranges are generally consistent with some large-scale 
flow determinations (e.g., \markcite{Willick:98,Baker:98}{Willick} \&
{Strauss} 1998; {Baker} {et~al.} 1998; \markcite{DNW:96}{Davis},
{Nusser}, \& {Willick} 1996) but not with the POTENT analyses, which
prefer $\beta\sim 1$ \markcite{Sigad:98}(e.g., {Sigad} {et~al.} 1998).
Of course, the bias may in general depend on scale, in which case our
small-scale result need not match the $\beta$ values measured using
flows on much larger scales.

Finally, we can combine the values of $\Omega_m$ and $\Omega_m/b^2$ to
obtain an estimate of the bias of the galaxy distribution on small
scales.  Our high-density models require biases $b=1.0$--$1.5$, 
while the low-density models are slightly anti-biased, $b=0.7$--$1.1$.
These ranges are consistent with the biases measured directly from the 
correlation functions of the simulations.

\subsection{Effects of Streaming Velocities}

Although our goal is to measure the particle distribution function
from redshift-space information alone, we must do this by considering
the relative motions of pairs of galaxies, for which we expect mean
streaming as well as thermal motions.  As defined in
Equation~\ref{eq:fv}, our model does not account for a non-zero first
moment of the velocity distribution of pairs of galaxies.  However,
the first moment will, in general, be non-negligible due to the mean
tendency of galaxies to approach each other, and it will contaminate a
measurement of the second moment.  On small scales in virialized
clusters, for example, the infall velocity approximately cancels the
Hubble expansion, and so its presence 
\eapjfig{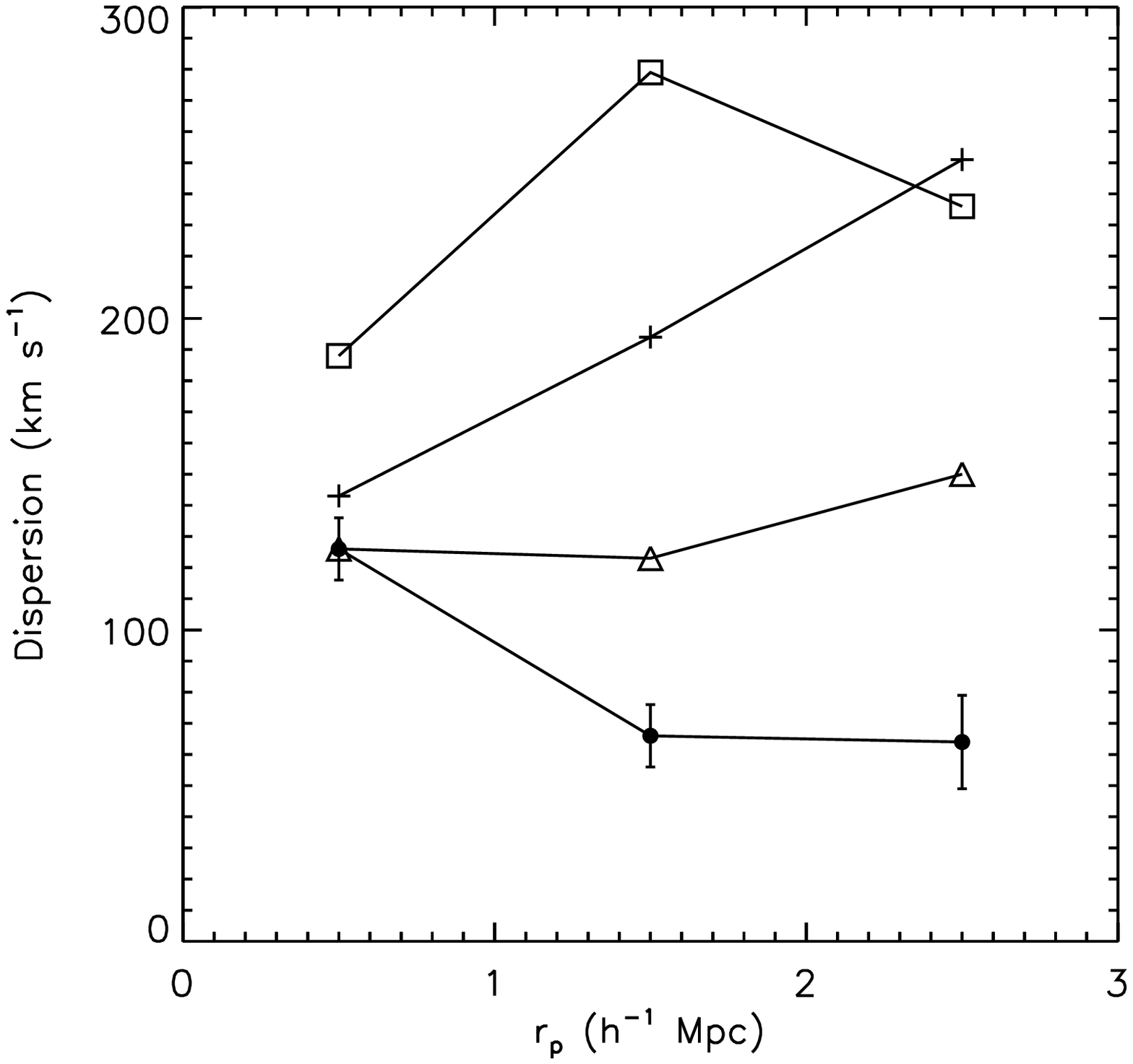}{Object-weighted velocity dispersion measured in independent 
  cylindrical shells of width $1\hMpc$.  The LCRS data are shown as
  filled circles with error bars.  Also shown are LCDM mass
  (squares) and halos drawn from our simulations (crosses) and from
  the Virgo simulation (triangles).
  \label{fig:scale}}{0.15in}
can affect our measurements on
$1\hMpc$ scales by of order $100\kms$.  \markcite{1998ApJ...503..502J}{Jing} \& {Boerner} (1998)
have shown that the effect of the streaming motions on the estimate of
the pairwise velocity dispersion can be dramatic, increasing
$\sigma_{12}$ from $\sim 400\kms$ to $580\kms$ at $1\hMpc$ separation.

The effects of the streaming motions can be incorporated into our
analysis by writing the distribution function in
Equation~\ref{eq:model} as
\begin{equation}
  \label{eq:fstream}
  f(v) = -\frac{1}{\sigma_1^\prime} \exp\left(-\frac{|v-\overline{v_1}|}
    {\sigma_1^\prime}\right),
\end{equation}
where $\overline{v_1}$ is the mean object-weighted streaming velocity,
which is a function of separation, and $\sigma_1^\prime$ is the second
moment of the streaming-corrected velocity distribution.  The form of
$\overline{v_1}$ is unknown but can be measured directly from $N$-body
simulations.

\begin{table*}
  \begin{center}
    \begin{tabular}{c|cc|c}
      $F$ & $\sigma_{1,\mathrm{LCRS}}^\prime$ & $\chi^2_\nu$ &
      $\sigma_{1,\mathrm{LCDM}}^\prime$ \\
      \tableline
       0  & 126$\pm$10 & 1.22 & 143 \\
      0.5 & 162$\pm$12 & 1.20 & 195 \\
       1  & 201$\pm$13 & 1.25 & 245 \\
      1.5 & 239$\pm$14 & 1.35 & 292 \\
    \end{tabular}
    \caption{Velocity widths with corrections for self-similar
      streaming motions applied.  Results are listed for the LCRS and
      for halos drawn from our LCDM simulation.}
    \label{tab:stream}
  \end{center}
\end{table*}

Our estimate of $\sigma_1$ with $\overline{v_1}=0$ will be smaller than
$\sigma_1^\prime$ because streaming motions tend to cause objects to
pile up at small velocity separations in redshift space.  However,
$\sigma_1$ has the advantage that it is a model-independent statistic,
relying only on the assumption of an exponential velocity
distribution.  The comparison of the data with $N$-body models is also
consistent; to the extent that the models describe the real universe,
the same streaming motions will be present in both the data and the
models, and will affect the estimates of $\sigma_1$ similarly.
Incorporating a non-zero $\overline{v_1}$ introduces model dependencies
into the measurement, and there is no guarantee that the infall
measured in the $N$-body simulations matches that of the real
universe.

For the application of the cosmic energy equation, it is in fact more
appropriate to use $\sigma_1$ rather than $\sigma_1^\prime$, because
contributions from both random thermal motions and mean streaming
motions are already included.  On the other hand, $\sigma_1^\prime$ is
a better measure of the truly thermal energy of the galaxy
distribution.  We can estimate it by using Equation~\ref{eq:fstream}
with an appropriate model for $\overline{v_1}$.  For the mean pairwise
velocity, the simple form
\begin{equation}
  \label{eq:v12dp}
  \overline{v_{12}}(r) = - \frac{F H_0 r}{1 + (r/r_0)^2},
\end{equation}
\markcite{Davis:83}(Davis \& Peebles 1983) is often used, where $F$ is a numerical factor,
typically $F=1$--$1.5$.  Another expression has been proposed more
recently by \markcite{Jusz:99}Juszkiewicz, Springel, \& Durrer (1999):
\begin{equation}
  \label{eq:v12pgf}
  \overline{v_{12}}(r) = -\frac{2}{3} f H_0 r
  \xibarbar(r) \left[ 1 + \alpha \xibarbar(r) \right],
\end{equation}
where $f\approx\Omega_m^{0.6}$, $\alpha \approx 1.2 - 0.65\gamma_0$
with $\gamma_0\equiv -d\ln\xi/d\ln r|_{\xi=1}$, and
\begin{equation}
  \label{eq:xibarbar}
  [1 + \xi(r)]\,\xibarbar(r) = \frac{3}{r^3} \int_0^r dx\, x^2 \xi(x).
\end{equation}
These two forms for $\overline{v_{12}}(r)$ are nearly equal at small
scales $r\lesssim 10\hMpc$ if we set $F=1.8\Omega_m^{0.6}$; note that
$F=1$ corresponds to streaming motions which just cancel the Hubble
expansion on small scales.

Table~\ref{tab:stream} shows that the streaming correction has a
substantial effect on the derived LCRS velocity width, with
$\sigma_1^\prime$ rising to $201\pm 13\kms$ for $F=1$ and $261\pm
15\kms$ for $F=1.8$.  The $\chi^2$ statistic worsens somewhat for
$F>1$.  The $N$-body models show similar behavior.  We caution,
however, that the streaming-corrected dispersions are model-dependent
and are not an appropriate measure of the single-particle dispersion
for use with the cosmic energy equation, which is defined in the
comoving frame of the universe.  This is in contrast to analyses of
the pair dispersion, where it is appropriate to use the cosmic virial
theorem, defined in the mean streaming frame.

\section{Conclusions}

Although the potential of small-scale cosmological velocities as a
cosmological probe has long been recognized, the application of
pair-weighted statistics is problematic.  We apply an extended version
of the more stable galaxy-weighted statistic of DMW to the Las
Campanas Redshift Survey.  We derive a one-dimensional rms velocity
for individual galaxies relative to their neighbors of $\sigma_1 =
126\pm 10\kms$ on scales $\sim 1\hMpc$.

Using this new statistic, we find that the observed velocities remain
quite cold relative to the predictions of high-$\Omega_m$ $N$-body
simulations.  Tilting the power spectrum to reduce the initial power
on small scales does little to resolve this discrepancy.  We have also
examined flat and open models with $\Omega_m=0.3$; these models
produce significantly lower dispersions than the high-density models.
Combining the LCRS data with the predictions based on halos in the
simulations, we measure consistent values $\Omega_m\sim 0.2$ for all
models, and we can rule out $\Omega_m=1$ with a high degree of
confidence.  Our result suggests that the extremely cold dispersion
measured in the vicinity of the Local Group (\markcite{Schlegel:94}Schlegel, Davis, \& Summers 1994;
\markcite{Governato:97}Governato {et~al.} 1997) might be a local anomaly, as currently popular
low-density models can reproduce the observed mean dispersion on
$1\hMpc$ scales.  On the other hand, at slightly larger separations,
we find evidence that all of the models may again be too hot relative
to the observations.

In the future, it will be extremely useful to apply our statistic to
upcoming redshift surveys, such as the Sloan Digital Sky and 2dF
surveys, which will contain enough galaxies to compute $\sigma_1$
precisely for different sub-samples of the galaxy population.  The
Deep Extragalactic Probe \markcite{DEEP:98}(DEEP; Davis \& Faber 1998) and other surveys at
high redshift will also provide a measure of the evolution of
$\sigma_1$, which can be used to place additional constraints on
cosmological parameters and the bias of the galaxy distribution.

\acknowledgments

J.~E.~B. acknowledges support from an NSF graduate fellowship.  This
work was supported in part by NSF grant AST95-28340.  HL acknowledges
support provided by NASA through Hubble Fellowship grant
\#HF-01110.01-98A awarded by the Space Telescope Science Institute,
which is operated by the Association of Universities for Research in
Astronomy, Inc., for NASA under contract NAS 5-26555.  We thank
C.~Frenk and A.~Benson for generously providing data from the Virgo
simulations, and we thank R.~Sheth and R.~Juszkiewicz for helpful
discussions.  We are also grateful to U.~Seljak and M.~Zaldarriaga for
making their CMBFAST code publicly available, and we thank
D.~Eisenstein for the HOP code.


\clearpage

\end{document}